\documentclass[aps,prl,twocolumn,groupedaddress,amsmath,amssymb,showpacs]{revtex4}
\usepackage{graphicx}
\usepackage{dcolumn}
\usepackage{bm}
\begin{document}

\title{Atomic Interferometer with Amplitude Gratings of Light and its
Applications to Atom Based Tests of the Equivalence Principle}



\author{Sebastian Fray$^1$, Cristina Alvarez Diez$^{1,2}$,
Theodor W. H\"ansch$^{1,2}$, and Martin Weitz$^3$}
\affiliation{ 1: Max-Planck-Institut f\"ur Quantenoptik, 85748 Garching, Germany\\ 2: Sektion Physik der
Universit\"at M\"unchen,80799 M\"unchen, Germany \\
3: Physikalisches Institut der Universit\"at T\"ubingen, 72076 T\"ubingen, Germany}
\date{\today}
\begin{abstract}
We have developed a matter wave interferometer based on the diffraction of atoms from effective absorption
gratings of light. In a setup with cold rubidium atoms in an atomic fountain the interferometer has been used to
carry out tests of the equivalence principle on an atomic basis. The gravitational acceleration of the two
isotopes $^{85}$Rb and $^{87}$Rb was compared, yielding a difference $\Delta \mathbf{g}/ \mathbf{g} =(1.2\pm 1.7)
\cdot 10^{-7}$. We also perform a differential free fall measurement of atoms in two different hyperfine states,
and obtained a result of $\Delta \mathbf{g}/ \mathbf{g} =(0.4\pm 1.2) \cdot 10^{-7}$.


\end{abstract}
\maketitle



Optical fields can be used to coherently split and recombine atomic de Broglie waves \cite{Berman,Dub85,Bord89}.
To date, atomic interferometry has allowed for impressive precision measurements of the earth's gravitation
\cite{Pet99}. Einsteins weak equivalence principle states that all bodies, regardless of their internal
composition, are affected by gravity in an universal way, i. e. in the absence of other forces they fall with the
same acceleration. For macroscopic (i.e. classical) objects, tests of the equivalence principle have been
performed since the early days of modern physics \cite{Su94}. Investigations of the equivalence principle on an
atomic basis have been proposed \cite{Lam93}, motivated by the quest to provide new tests of theories which merge
quantum mechanics and relativity.

Here we report of an experiment demonstrating a test of the equivalence principle with quantum probe particles
based on atom interferometry. Our experiment uses the two stable isotopes of the rubidium atom with nearby optical
transition wavelength. For our measurements, we have developed a light pulse atom interferometer based on the
diffraction of atoms from standing optical waves acting as effective absorption gratings. By comparing the free
fall of the two distinct isotopes $^{85}$Rb and $^{87}$Rb, we have measured their differential acceleration in the
earth's gravitational field to a relative accuracy of 1.7$\cdot$10$^{-7}$. Further, we have tested for a variation
of the measured free fall acceleration as a function of relative orientation of nuclear to electron spin to a
differential accuracy of 1.2$\cdot$10$^{-7}$ by comparing interference patterns measured with $^{85}$Rb atoms
prepared in two different hyperfine ground states. Within the quoted uncertainties, the results in both of our
measurements are consistent with an isotope and internal state independent gravitational acceleration. We expect
that technical improvements will in future allow for very critical tests of the equivalence principle on an atomic
basis.

Before proceeding, we note that the first atom based tests of general relativity are the famous Pound-Rebka
experiments, which are sensitive to the gravitational redshift. For a comparison of different tests of the
equivalence principle see \cite{Nordtvedt}. To date, this tiny shift (of order $\Delta h g/c^2$) has been verified
to an accuracy of 10$^{-4}$ \cite{Vess80}. We also wish to point out that theoretical work has discussed the
possibility of a spin-gravitational coupling \cite{Lam97}. Such theories have so far only been tested with
macroscopic matter \cite{Hsi89}.


When comparing the free fall of different atomic species or atoms in different internal states, it seems
advantageous to use techniques which are only weakly dependent on a specific internal atomic structure. In the to
date most absolute atomic interferometric measurement of gravitation \cite{Pet99}, beam splitters based on
off-resonant Raman transitions were used which change both the internal and external degrees of freedom. Another
class of atom interferometers only involve interference of the external degrees of freedom, as have e.g. been
realized with off-resonant standing waves acting as phase gratings \cite{Rasel}.

    \begin{figure}[b]
    \includegraphics[width=8.5cm]{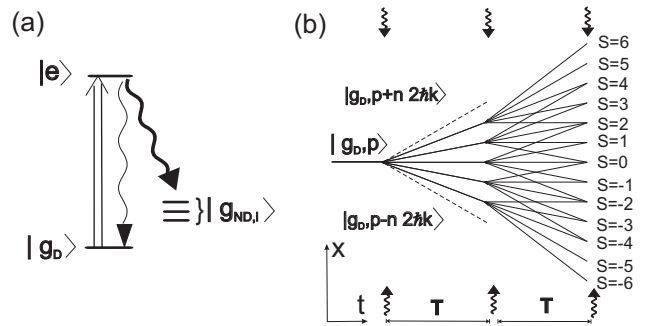}%
     \caption{\label{fig_paths}(a) Simplified level scheme
     and (b) recoil diagram of an atomic interferometer realized
    with three effective absorption gratings of light.}
    \end{figure}

Our scheme for atomic interferometry is shown in Fig. 1. The atomic beam splitters are realized with pulsed
standing waves tuned resonantly to an open transition from a ground state $|g_{D}\rangle$ to a spontaneously
decaying excited state $|e\rangle$. After the pulses only atoms in state $|g_{D}\rangle$ are detected. For an
incident atom distribution, the standing wave fields will lead to a spatially dependent pumping, and atoms passing
near the antinodes will preferentially be ''removed'' into states $|g_{ND,i}\rangle$, which are not detected,
while atoms near the nodes will remain in state $|g_{D}\rangle$. The transient standing wave thus acts as an
effective absorption grating of spatial periodicity $\lambda/2$, on which an incident plane atom wave
$|g_{D},\vec{p}\rangle$ with internal state $|g_{D}\rangle$ and momentum $\vec{p}$ is diffracted into a series of
components ..., $|g_{D},\vec{p}-2\hbar\vec{k}\rangle,|g_{D},\vec
{p}\rangle,|g_{D},\vec{p}+2\hbar\vec{k}\rangle,...$ . For an increased pulse energy, the open fraction of the
effective grating lessens, which results in a larger number of generated paths.

Since only resonant light is used, we do not expect phase shifts between adjacent paths due to the ac Stark
effect. While diffraction from a single effective optical absorption grating has been demonstrated \cite{Pre97},
the extension to an atomic interferometer is not obvious since the spontaneous emission occurring during the beam
splitting process may a priori affect the coherence, which would destroy the observed interference contrast. The
operation of our atom interferometer demonstrates that this is not the case, which intuitively can be seen by
noting that the use of an open optical transition ensures that, assuming a favourable branching ratio, which-path
information is only present for atoms that have undergone an optical pumping event. Our experiment relies on a
rubidium atomic fountain, in which a temporal sequence of three resonant optical standing wave pulses is applied
to coherently split and redirect the wavepackets, and finally readout the atomic interference pattern. The
diameter of the pulsed optical beams is sufficiently large that each pulse affects the whole atomic cloud.


In our experiment, the ground state $|g_{D}\rangle$ was chosen to be a magnetic field insensitive $(m_{F}=0)$
Zeeman sublevel of a rubidium ground state hyperfine component (i.e. F=2 or F=3 for $^{85}Rb$, or F=1 for
$^{87}Rb$). The standing wave is $\sigma^{+}-$polarized and tuned resonantly to a hyperfine components of the
rubidium D1-line. An atom interferometer is realized with a sequence of three such optical pulses. The first
standing wave pulse at t=0 splits an atomic wavepacket into several distinct paths. At a time t=T second pulse is
applied, which again leads to a diffraction and coherently splits up the paths.
 At time t=2T, several paths spatially overlap in a series of families of wavepackets. We expect, that the wave
nature leads to a spatial atomic interference structure. To read out this periodic fringe pattern, we again apply
a resonant optical pulse tuned to the open transition. The periodic pattern could now be read out by scanning the
position of this third grating and monitoring the number of transmitted atoms in state $|g_{D}\rangle$. For
technical reasons, we actually leave the position of the optical grating constant and instead vary the pulse
spacing T, which due to the earth's gravitational acceleration also allows us to observe the fringe pattern.

We shall now outline the calculation of the interference pattern (see also \cite{Dub85}). The Hamiltonian of the
system is assumed to be $H= \vec{p}^2/2m+( \hbar \omega_{eg}-i\hbar \Gamma /2) |e\rangle \langle e|
    - e\vec{r} \cdot \vec{E}$
where $\hbar\omega_{ge}$ denotes the atomic energy spacing between level $|g_{D}\rangle$ and $|e\rangle$. The
relaxation of the excited state is accounted for by introducing a non-Hermitian decay term
-$\frac{i\hbar\Gamma}{2}.$ This neglects the decay into the levels $|g_{D}\rangle$. In the case of negligible
branching ratio of the decay into $|g_{D}\rangle$
 the above Hamiltonian would allow for an exact description of the observed fringe pattern.
Otherwise, we expect an additional incoherent background to the fringes. The standing wave electric field is
assumed to be $\vec{E}$ = $\vec{E}_{0}\cos(\vec{k}\vec{r}-\omega t)+\vec{E}_{0}\cos(-\vec {k}\vec{r}-\omega t).$

We shall first consider the atomic response to a single pulse of the standing wave field. Assume that the pulse
length $\tau$ is so short that we can neglect the phase evolution due to the kinetic energy term
 (Raman-Nath regime \cite{Berman}). A plane atomic wave with initial internal and external states $|g_{D},\vec{p}\rangle$
is then diffracted into the coherent superposition of paths
\begin{equation*}
|\Psi \rangle=\sum^{\infty}_{n=-\infty} \exp \Big( -\frac{\Omega^2}{\Gamma}\tau_1 \Big) ~I_n
\Big(-\frac{\Omega^2}{\Gamma}\tau_1 \Big) ~|g,\vec{p}+2n\hbar\vec{k}\rangle
\end{equation*}
where $\Omega=e\vec{E}_{0}\langle g|\vec{r}|e \rangle\hbar$ denotes the Rabi frequency and I$_{n}(x)$ the modified
Bessel function. Unlike the "usual" Bessel functions J$_{n}(x)$ , which describe diffraction by an off-resonant
standing wave, the functions I$_{n}(x)$ do not oscillate with x. Because of this property optical absorption
gratings are promising tools for future multiple beam atom interferometers with large number of paths
\cite{Weit96, Hin97}. The approximate atom loss equals the ''open ratio'' of the effective absorption grating, and
is of order 1/N for an N-path beam splitter.

Using the above Hamiltonian one can calculate the momentum picture wavefunction directly before the third
(readout) pulse. We are interested in the atomic response for a broad velocity distribution, for which the
velocity width along $\vec{k}$ is large compared to 1/(k$\cdot T$). In the expression for the spatial density
$\rho_{D}(x)=\int g(p)\cdot\langle x|\psi\rangle \langle\psi|x\rangle dp$ of atoms in state $|g_{D}\rangle$, we
can solve the integral and simplify the resulting expression to
$\rho(x)=\sum^{\infty}_{s=-\infty}||\Psi_s(x)\rangle|^2$, where
\begin{eqnarray} |\Psi_{s}\rangle&&=\sum^{\infty}_{n=-\infty}~I_n \Big(-\frac{\Omega^2}{\Gamma}\tau_1 \Big)
~I_{s-2n} \Big(-\frac{\Omega^2}{\Gamma}\tau_2\Big) \nonumber\\
 &&\times\quad e^{ -\frac{\Omega^2}{\Gamma}(\tau_1+\tau_2)}
 \times e^{i(sn-n^2) 2\omega_{r}T} e^{-in2\vec{k}\vec{x}}\nonumber
\end{eqnarray}
Here, $s=2n+l$ and $\omega_{r}=2\hbar k^{2}/m$ denotes the recoil energy of a two-photon transition in frequency
units. In the presence of a gravitational field, we obtain an additional phase term -$in$ $\vec{k}\vec{g}
T^{2\text{ \ \ }}$in the exponential factor. While the expected density modulation is
near sinusoidal in the unsaturated case (i.e. $\frac{\Omega^{2}}{\Gamma}%
\tau_{i}\lesssim1)$, the fringe pattern sharpens to a Airy-function like multiple-beam pattern for larger pulse
energies \cite{Weit97}.

Our experimental setup is based on a ultrahigh vacuum chamber, in which $2\cdot 10^9$ atoms ($^{85}$Rb or
$^{87}$Rb respectively) are initially captured in a magnetooptical trap
and accelerated 
by an upwards traveling optical molasses. 
The temperature of the atoms at this point is 6 $\mu$K. After the launch, a homogeneous 50 mG magnetic bias field
oriented parallel to the optical interferometer beams is activated. While the atoms travels on a fountain-type
ballistic trajectory, a series of beam splitting pulses is applied from the optical interferometer beams. Before
and after the interferometric pulses we select the $m_F=0$ component of the corresponding hyperfine state by an
appropriate sequence of microwave $\pi$-pulses and a resonant optical pulse removing residual population. The
atomic signal is then measured with a FM-fluorescence detection method, for which the atoms were irradiated with a
resonant modulated optical beam. The cold atoms convert the frequency modulation into an amplitude modulated
fluorescence, which is phase-sensitively detected. This method was developed to selectively detect the signal of
cold atoms in the presence of a thermal rubidium background vapour. Further improvements on the signal to noise
ratio would be possible with a differentially pumped vacuum chamber.


The optical interferometer pulse sequence is generated by an injection locked master-slave diode laser system
operating near 795 nm and a passage through two acoustooptic modulators. The light is coupled into a single mode
optical fiber, expanded to a 3 cm Gaussian beam diameter and coupled into the vacuum chamber. After the
transmitting the chamber, the beam is retroreflected with a mirror. The generated (pulsed) optical standing wave
irradiating the cold atoms is oriented vertically, and has an intensity of 3 mW/cm$^2$ per direction. The
retroreflection mirror is passively vibration isolated by
 suspending the mirror on a 2.5 m long ribbon string.

The atom interferometer as shown in Fig. 1 is realized applying a temporal sequence of three resonant standing
wave pulses to the cold atoms. The length of an optical beam splitting pulse is 200 ns. We observe fringe patterns
with good signal to noise up to interrogation times near T=40 ms. Fig. 2 shows typical atomic signals as a
function of pulse spacing T between the optical beam splitting pulses. The observed fringe contrast is 20 per cent
for small interrogation times T and reduces to some 9 per cent for the shown fringe patterns with high
gravitational sensitivity recorded near T=40 ms. This loss of fringe contrast for longer coherence times is
attributed to residual mechanical vibrations of the interferometer beams retroreflecting mirror. The widths of the
observed principal maxima is between 0.4$\cdot2\pi$ and 0.5$\cdot2\pi$. The former value was reached when using
longer optical pulse lengths, which however lessened the signal amplitude. We attribute this mainly to residual
reflection of the optical vacuum windows, which cause an intensity imbalance between forward and backward running
wave of some 2 per cent. The imperfect window transmission of our present apparatus is attributed to contamination
with adsorbed rubidium from the background vapour. Applying a white light desorption method \cite{And01} lowers
the disturbing reflection from 6 to the above quoted value of 2$\%$. We were however not able to completely remove
the contamination. It is anticipated, that the residual intensity imbalance also reduces the fringe contrast from
an theoretical value near 100$\%$ for the used transition to the experimentally observed value at small drift
times T.

    \begin{figure}[t]
    \includegraphics[width=8.5cm]{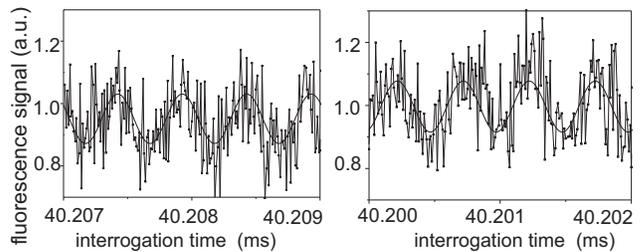}
    \caption{\label{fig_fringe} Typical interference signals
    as a function of pulse spacing T varied in narrow
    regions near T=40 ms.
    The signals were recorded
    for (a) $^{85}$Rb atoms with the optical field tuned to the
    F=2$\rightarrow$F'=3 component near the 1195th revival 
    and for (b) $^{87}$Rb atoms using F=1$\rightarrow$F'=2 
    component of the rubidium D1-line near the 1167th revival, respectively.
    In the presence of the earth's gravitational field an observation of the interference pattern
    is possible by a variation of the interrogation times.
    Each data point corresponds to the
    signal recorded in a single fountain launch. The solid lines are
    fits to sinusoidal functions.
    }
    \end{figure}

    \begin{figure}[b]
    \includegraphics[height=2.0cm]{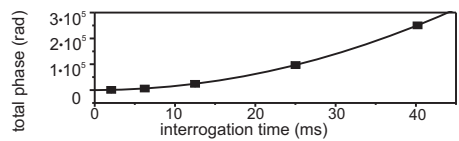}%
    \caption{\label{fig_qphase}
    Total atom phase between adjacent paths of the $^{85}$Rb interferometer as a function
    of interrogation time T. The solid line represents a parabolic fit. The used atomic levels were as in Fig. 2a.
    }
    \end{figure}

To determine the total phase shift between the interfering paths induced by the earth's gravitational field, we
have analyzed fringe patterns recorded for different pulse spacings T, which avoids any phase ambiguity caused by
the periodic character of the pattern.  Figure \ref{fig_qphase} shows typical recorded data (dots) along with a
fit to the expected parabolic curve (solid), while the gravitational acceleration g was left as a free parameter.
From the fringe patterns recorded at T = 40 ms interrogation time, we at present can determine the earth's
gravitational acceleration to a statistical accuracy of 8$\cdot10^{-8}$ in 6 hours of data acquisition. It is
anticipated that the implementation of an active vibration isolation stage, as used in ref. \cite{Pet99}, would
allow for longer interrogation times and yield a comparable statistical accuracy for measurements of the
gravitational phase.


We have compared the gravitational acceleration experienced by $^{85}$Rb atoms in the $F=2,m_{F}=0$ hyperfine
ground state with that of $^{87}$Rb atoms in $F=1,m_{F}=0.$ For this atom based test of the equivalence principle,
we recorded series of fringe patterns with T$\simeq40$ ms for the two different isotopes. The data were recorded
in three measurement sessions, each corresponding to the data recorded during one day. Fringe patterns have for
both measurements already been shown in Fig. 2. Fig. 4a shows the measured averaged values of
$(\mathbf{g}_{^{85}Rb}-\mathbf{g}_{^{87}Rb})/\mathbf{g}_{^{85}Rb}$ in the three measurement sessions along with a
vertical line representing the total average. This final value for the difference is
$(\mathbf{g}_{^{85}Rb}-\mathbf{g}_{^{87}Rb})/\mathbf{g}_{^{85}Rb}=(1.2\pm 1.7)\cdot10^{-7}$. The quoted
statistical error here equals the (estimated) total uncertainty. In our differential measurement, systematic
errors due to misalignment of the beams, wave front curvature, and Coriolis forces largely cancel, and at the
present level of accuracy clearly can be neglected. Moreover, as all paths of our atom interferometer are in the
same internal state, systematic effects due to the second order Zeeman shift only occur in the presence of a
magnetic field gradient. The estimated systematic uncertainty due to field inhomogenities (8 mG/cm) is for our
present apparatus estimated to be $5\cdot10^{-11}$, which is also clearly negligible. Within the quoted
uncertainties, our above given value agrees well with the expected result of an identical gravitational
acceleration of the two rubidium isotopes.

    \begin{figure}[t]
     \includegraphics[width=8.5cm]{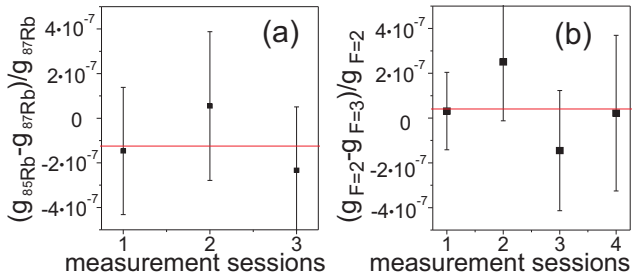}
     \caption{\label{fig_vergl} Measured difference of the gravitational acceleration in individual
    measurement sessions (dots) and total average (vertical line) for (a) the two different isotopes $^{85}Rb$ and
    $^{87}Rb$ and (b) for atoms ($^{85}Rb$) in the different hyperfine ground states $F=2$ and $F=3$.}
    \end{figure}

We have furthermore compared the gravitational acceleration experienced by atoms ($^{85}$Rb) in the two different
hyperfine ground states $F=2,m_{F}=0$ and $F=3,m_{F}=0$. For the latter measurement, the optical interferometer
beams were tuned to the F=3$\rightarrow$F'=3 hyperfine component of the D1 line. Our experimental results for the
measured relative differential gravitational acceleration is summarized in Fig. 4b. Our averaged, final result
here is $(\mathbf{g}_{F=3}-\mathbf{g}_{F=2})/\mathbf{g}_{F=2}=(0.4\pm1.2)\cdot 10^{-7}$. Within the quoted
uncertainty, we do not observe a difference in the earth's gravitational acceleration for atoms in the two
hyperfine states. Our atom interferometer in a very natural way allows for a comparison of results that are based
on different atomic states. An important question for future theoretical work is to investigate to what extend
such a measurement of gravitational acceleration of atoms in two different internal states is complimentary to
atomic clock experiments of the Pound-Rebka type.

To conclude, we have demonstrated an atom interferometer based on pulsed, effective absorption gratings of light.
We have applied the interferometer to demonstrate novel atom based test of the equivalence principle.

For the future, improvements in wavefront quality and a better vibrational isolation can yield longer coherence
times, and furthermore allow for multiple beam interference signals with sharp principal maxima where in contrast
to earlier schemes the path number is not limited by the internal atomic structure \cite{Weit96, Hin97}.
Furthermore, we anticipate that atom interferometry will allow for very critical test of the equivalence principle
on an atomic basis.

We acknowledge discussions with C. L\"ammerzahl, C. J. Bord$\acute{e}$ and R. Chiao. This work was supported in
parts by the Deutsche Forschungsgemeinschaft.


\begin{thebibliography}{99}
\bibitem{Berman}See, e.g.: P. Berman, Atom Interferometry (Academic
Press, San Diego, 1997).
\bibitem{Dub85} V. P.Chebotayev et al., J. Opt. Soc. Am. B \textbf{2}, 1791 (1985).
\bibitem{Bord89} C. J. Bord$\acute{e}$, Phys. Lett. A \textbf{140}, 10 (1989).
\bibitem{Pet99} A. Peters, C. Keng Yeow, and S. Chu, Nature \textbf{400}, 849 (1999).
\bibitem{Su94} Y. Su et al., Phys. Rev. D \textbf{50}, 3614 (1994)
\bibitem{Lam93} J.Audretsch, U. Bleyer, and C. L\"ammerzahl,
Phys. Rev. A \textbf{47}, 4632 (1993); L. Viola and R. Onofrio, Phys. Rev. D \textbf{55}, 455 (1997)
\bibitem{Nordtvedt} K. Nordtvedt, arxiv:gr-qc/0212044 (2004)
\bibitem{Vess80} R. F.C. Vessot et al. Phys. Rev. Lett., \textbf{45}, 2081 (1980)
\bibitem{Lam97} C. L\"ammerzahl, Proc. of the Intern. School of Cosmology and Gravitation, Course XV,
    edited by P. G. Bergmann et al.,
    (World Scientific, Singapore 1998) and ref. therein.
\bibitem{Hsi89} C.-H. Hsieh  et al., Mod. Phys. Lett. A \textbf{4}, 1597 (1989).
\bibitem{Rasel} E.M. Rasel et al., Phys. Rev. Lett. \textbf{75}, 2633 (1995);
D. M. Giltner, R. W. McGowan, and S. A. Lee, ibid. \textbf{75}, 2638 (1995); S. B. Cahn et al., ibid. \textbf{79},
784 (1997); S. Gupta et al., ibid. \textbf{89}, 140401 (2002).
\bibitem{Pre97} A. P. Chu, K. S. Johnson and M. G. Prentiss, Opt. Comm. \textbf{134}, 105
(1997); M. K. Oberthaler et al., Phys. Rev. Lett. \textbf{77}, 4980 (1996); A. Turlapov, A. Tonyushkin, and T.
Sleator, Phys. Rev. A \textbf{68}, 023408 (2003).
\bibitem{Weit96} M. Weitz, T. Heupel, T. W. H\"ansch, Phys. Rev. Lett. \textbf{77}, 2356 (1996).
\bibitem{Hin97} H. Hinderth\"ur et al., Phys. Rev. A \textbf{56}, 2085 (1997).
\bibitem{Weit97} M. Weitz, T. Heupel, T. W. H\"ansch, Appl. Phys. B \textbf{65}, 713 (1997).
\bibitem{And01} B. Anderson et al., Phys. Rev. A \textbf{63}, 023404 (2001).


\end{thebibliography}
\end{document}